\documentclass[showpacs,preprintnumbers,english,amsmath,amssymb]{revtex4}
\usepackage[T1]{fontenc}
\usepackage[latin1]{inputenc}
\usepackage{graphicx}
\usepackage{amssymb}
\usepackage{verbatim}
\usepackage{epic,eepic}
\usepackage{babel}
\usepackage{color}

\def\Journal#1#2#3#4{{#1} {#2}, #3 (#4)}

\def\ARNPS{{\em Ann. Rev. Nucl. Part. Sci.}}

\def\EPJA{{\em Eur. Phys. J. A}}
\def\IJMPE{{\em International Journal of Modern Physics E}}
\def\JCHP{{\em J. Chem. Phys.}}

\def\JPG{{\em J. Phys. G: Nucl. Part. Phys.}}

\def\NPA{{\em Nucl. Phys.} A}

\def\PRC{{\em Phys. Rev.} C}

\def\PRL{\em Phys. Rev. Lett.}

\def\ZPA{{\em Z. Phys. A.}}

\begin{document}

%\title{Pastina formation in low density nucleonic matter -- a mechanism for ternary fission}
\title{Nucleation and cluster formation in low-density nucleonic matter: A mechanism for ternary fission}

\author{S. Wuenschel$^{1}$, H. Zheng$^{1}$, K. Hagel$^{1}$, B. Meyer$^{2}$, M. Barbui$^{1}$, E.J. Kim$^{1, 3}$, G. R\"opke$^{4}$ and J. B. Natowitz$^{1}$}
\affiliation{
$^{1}$Cyclotron Institute, Texas A\&M University, College Station, TX 77843, USA;\\
$^{2}$Department of Physics and Astronomy, Clemson University, Clemson, SC 29634, USA;\\
$^{3}$Division of Science Education, Chonbuk National University, Jeonju 561-756, Korea;\\
$^{4}$University of Rostock,  FB Physik, Rostock, Germany.}

%\date{September 8, 2010}

\begin{abstract}
Ternary fission yields in the reaction $^{241}$Pu(n$_{th}$, f) are calculated using a new model which assumes a nucleation-time moderated chemical equilibrium in the low density matter which constitutes the neck region of the scissioning system. The temperature, density, proton fraction and fission time required to fit the experimental data are derived and discussed. A reasonably good fit to the experimental data is obtained. This model provides a natural explanation for the observed yields of heavier isotopes relative to those of the lighter isotopes, the observation of low proton yields relative to $^{2}$H and $^{3}$H yields and the non-observation of  $^{3}$He, all features  which are shared by similar thermal neutron induced and spontaneous fissioning systems.  
\end{abstract}


\maketitle

In the fission of a heavy nucleus, approximately $0.3\%$ of binary fission decays observed in thermal neutron induced or spontaneous events are accompanied by emission of an energetic light particle or fragment in a direction perpendicular to the axis defined by the separating massive fragments \cite{wagemans, book1, halpern, theobald, mehta, serot, heeg, koester, koester1, tsekhanovich}.  The discovery of ``long range alpha accompanied fission'' in 1946 \cite{san} and later experiments demonstrated the existence of a large range of light isotopes with characteristics  similar to those of the long range alphas \cite{heeg, koester, koester1, tsekhanovich}.  Considerable theoretical and experimental effort has been directed towards understanding this type of ternary fission \cite{wagemans, book1, halpern, theobald, mehta, serot, heeg, koester, koester1, tsekhanovich, valskii, schafer, rubchenya, lestone}. 

Although fission processes inherently involve an important collective dynamics,  the ternary isotope data indicate that statistical considerations also play a major role in the determination of the observed yields.  As a result, models which are primarily based on dynamic considerations \cite{wagemans,  halpern, rubchenya}, on statistical considerations \cite{wagemans, valskii} and on both \cite{wagemans, lestone} have been employed in attempts  to reproduce the observed yields. These previous efforts have treated the problem by emphasizing the dominance of one of the following: neck instabilities and rupture \cite{rubchenya}, barrier penetration of pre-formed clusters \cite{lestone} or formation of fragments from interacting nucleons in  the fission neck region \cite{valskii}. However, these models have had only limited success and have been unable to explain such key experimental results as the high yield of scission tritons relative to scission protons \cite{wagemans, book1, schubert} and the non-observation of $^3$He \cite{wagemans, book1, koester, koester1}. Recently, Lestone proposed a model in which a statistical evaporation  of the ternary particle is moderated by time dependent emission barriers which evolve as the fissioning nucleus approaches the scission point \cite{lestone}. Parameterizing the neck radius, the range of the nuclear force, temperature, time, and emission barrier height provided a good reproduction of isotopic yields for $Z\le 6$ and reasonable predictions for $Z>6$.

In this work, we show that the observed ternary fission yields in the reactions of $^{241}$Pu with thermal neutrons can be interpreted as reflecting nucleation \cite{clouet} and cluster formation probabilities in the low density neck between the two large fragments. This leads to achievement of a time-modulated approach to nuclear statistical equilibrium that, in addition to being in good agreement with the observed yields, offers a natural explanation for the observation high triton to proton yield ratio, and observational absence of $^{3}$He. This cluster production in the low density neck region might be viewed as the formation of a nuclear pastina, the mesoscopic analog of the nuclear pasta postulated in the skins of neutron stars. 

The experimental results of Koester {\it et al.} provide the most comprehensive data available for ternary fission yields \cite{koester, koester1}. For this study we focus on the data for the  ${}^{241}$Pu(n$_{th}$, f) reaction \cite{koester1}. These experiment data include measured yields per fission event for 42 isotopes.  In addition, 17 upper limits are also reported for yields of other isotopes. The relative yields are normalized to an assigned value of 10,000 for the $^{4}$He isotope as is commonly done in ternary fission papers \cite{heeg, koester, koester1, tsekhanovich}.  

\begin{figure}
\centering
\includegraphics[width=0.5\columnwidth]{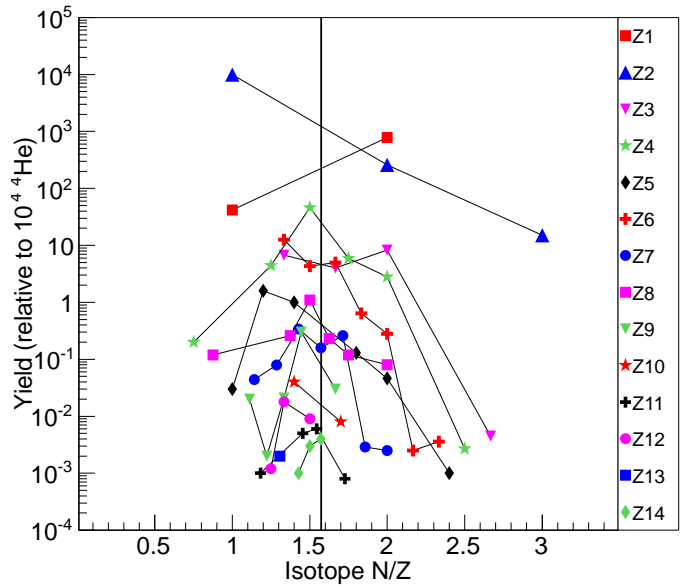}
\caption[]{(Color online) Relative yields of ternary cluster isotopes emitted in the fission of $^{242}$Pu. All yields are normalized to an assigned value of 10000 for $^{4}$He. Symbols represent the yields. Lines are added to guide the eye. The vertical line indicates N/Z of the $^{242}$Pu, 1.574. } \label{Fig1}
\end{figure}
\begin{figure}
\centering
\includegraphics[width=0.5\columnwidth]{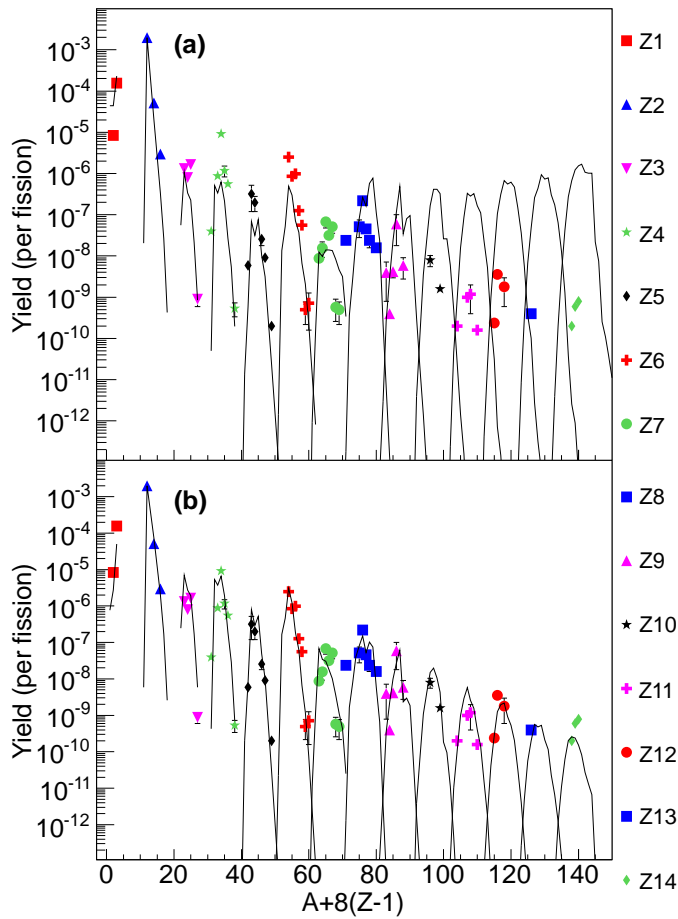}
\caption[]{(Color online) Yield per fission as a function of mass(A) and charge(Z) of products. Solid points represent $^{241}Pu(n_{th},f)$ experimental yields from Koester {\it et al} \cite{koester1}.  Lines are theoretical predictions from NSE calculation \cite{heeg}. NSE parameters are $T= 1.4 MeV$, $\rho= 4\times 10^{-4} fm^{-3}$, and $Y_p= 0.34$.  a) NSE calculation only. $M^2$ fit metric = 4.28. b)  NSE calculation with nucleation. Nucleation parameters are time $= 6400 fm/c$ and $A_c = 5.4$. Fit metric $= 1.18$. } \label{Fig2}
\end{figure}

In Figure \ref{Fig1} we present a chart of relative yields of the isotopes observed as a function of the isotopic N/Z ratio. Upper limits are included. This plot emphasizes that the $^{4}$He yield is dominant, accounting for about $50\%$ of the total ejected ternary particle mass. The vertical line in the plot represents N/Z = 1.574, that of the $^{242}$Pu fissioning system. We note that yields of heavier elements cluster near this line with peak yields slightly below the line. In Figure \ref{Fig2} a and b the absolute experimental yields for the ternary particles are plotted. Upper limits are not included. Both experimental and theoretical data are converted absolute yields using the ternary to binary ratio for $^{242}$Pu fission events \cite{wagemans1}. To visually separate yields for different elements and isotopes we have plotted the yields as a function of  the parameter 8(Z-1)+A suggested by Lestone \cite{lestone}.   Here Z is atomic number and A is mass number. 

For our initial approach to modeling the yield data, we employed a nuclear statistical equilibrium code (NSE) built on top of libnuceq \cite{meyer} to determine the relative yields of the constituent species.  The key assumption of nuclear statistical equilibrium is that the chemical potential $\mu(Z,A)$  is governed by the equation
\begin{equation}
\mu(Z,A) = Z\mu_p + (A-Z)\mu_n \label{eq1}
\end{equation}
where $\mu_p$ and $\mu_n$ are the proton and neutron chemical potentials, respectively.  The yields follow from the relationship given in Eq \ref{eq2}.
\begin{equation}
\mu(Z, A) = m(Z, A)c^2 + kT \ln \Big(\frac{\rho N_A Y(Z, A)}{G(Z, A)}\Big[\frac{h^2}{2\pi m(Z, A)kT}\Big]^{3/2}\Big) \label{eq2}
\end{equation}

In this equation, $m(Z, A)$ is the mass, $k$ is Boltzmann's constant, $T$ is Temperature, $\rho$ is the density, $N_A$ is Avogadro's number, $Y(Z, A)$ is the yield and $G(Z, A)$ is the nuclear partition function. The partition function for a given nuclear species incorporates excited states as multiples of the ground state.  For nuclei above $Z =7$ the modified partition functions of Rauscher {\it et al.} \cite{rauscher}, determined for temperatures up to $1.4 MeV$,  have been employed.  Experimental binding energies were obtained from the JINA astrophysical database \cite{w1}.  The input parameters of the NSE calculation are temperature, density, and proton fraction.

Several different attempts to evaluate the temperatures appropriate to thermal neutron induced ternary fission have led  to temperatures in the range of $1.0$ to $1.4 MeV$ \cite{valskii, andronenko}. At such low temperatures, cluster formation is expected to occur at densities well below normal nuclear densities \cite{meyer, ropke}. For the $^{242}$Pu compound nucleus the proton fraction, $Y_p$, is $0.388$. While the proton fraction of the fissioning system should be close that of the compound system, various theoretical models  suggest that the region between the separating fragments, which dominates the production of the ternary particles, will be neutron enriched \cite{sobotka}. 

After surveying results for a wide variety of temperature, density, and proton fraction values we adopted a multi-parameter minimization technique to simultaneously fit the available experimental data. The fit metric used used is that of Lestone \cite{lestone}, defined by 
\begin{equation}
M^2 = \sum_j \{ \ln [P_{TF}^{exp}(Z_j, A_j)] - \ln [P_{TF}(Z_j, A_j)] \}^2/n, \label{eq3}
\end{equation}
where $P_{TF}$ are the calculated ternary fission probabilities, $P_{TF}^{exp}$ are the corresponding absolute experimental emission probabilities, and $n$ is the number of fitted experimental data points. The exponential of $M$ is a measure of the typical relative difference between the model calculations and the experimental data. For $M\approx1$  the average relative discrepancy between model and experiment would be a factor of $\sim 3$. 

While this approach produced reasonable fits for the lighter isotope yields A$\le15$, it greatly overestimated the yields for heavier isotopes. The results of one such calculation are presented in Figure \ref{Fig2}a. The choice of parameters used there, $T= 1.4 MeV$, $\rho= 4\times 10^{-4} fm^{-3}$, and $Y_p= 0.34$, is based on extensions of the fitting model described below and plotted in Figure \ref{Fig2}b.

The use of a NSE code to model fragment production encounters two major conceptual issues.  First, the NSE code has an infinitely large source of free nucleons from which it can generate clusters while the number of nucleons in the fissioning nucleus is limited. The nucleus is finite and average multiplicities of ternary ejectiles are of the order of $0.003$ per fission events. Thus, no equilibrium distribution of ternary isotopes such as might be established in infinite nuclear matter is expected to exist in a single nucleus. The calculation carried out here should be interpreted as establishing the relative probabilities for the formation of the observed (and unobserved) isotopes. Secondly, coalescence of nucleons into clusters is a dynamic process requiring time while the fissioning system exists for a limited time span.  With the NSE code, we have not constrained the size of the system.  However, since the observed ternary clusters require not more than $\sim 15\%$ of the nucleons, the first consideration appears not to be a serious limitation. On the other hand, we find that the dynamic time limit is important and introduce it using a nucleation model approach to the cluster formation. While developed primarily in a chemical kinetics context \cite{clouet, wilemski}, nucleation models have been proposed for nuclear processes \cite{demo, schmelzer}. Such approaches have much in common with thermal coalescence approaches previously applied to clustering in low density nuclear systems \cite{mekjian, hagel} but explicitly incorporate consideration of cluster formation rates. 

Nucleation has been widely studied in chemical systems undergoing phase changes \cite{clouet, wilemski}. Fragments are formed under the constraints of surface tension and chemical potentials.  Clusters grow or diminish through capturing or releasing  single nucleons.  The probability of larger mass changes is expected to be orders of magnitude lower and thus is generally neglected \cite{schafer}. For application to nucleation in nuclear matter Demo and Kozisek have proposed a single component nucleation model which allows derivation of a relatively simple analytical expression for the yield distribution as a function of normalized time $\tau = \frac{3.967 c\rho}{A_c^{2/3}\sqrt{T}}t$, where $\rho$ is density, $A_c$ is the critical cluster size, $T$ is temperature, and $t$ is time \cite{san}. That expression is 
\begin{equation}
Y(A, \tau) = \frac{1}{2}\rho \exp\Big[-\frac{G(A)}{T}\Big] erfc \Big[B(T, \sigma) \frac{[(A/A_c)^{1/3}-1]+(1-A_c^{-1/3})\exp(-\tau)}{\sqrt{1-\exp(-2\tau)}}\Big] \label{eq4}
\end{equation}
where the $\rho \exp[-\frac{G(A)}{T}]$ term representing the equilibrium concentration of the species of mass $A$ is modulated by a complementary error function term  which depends upon the parameters $B(T, \sigma)$ and $A_c$ where   
\begin{equation}
B(T, \sigma) = 2R_0 \Big(\frac{\pi \sigma}{T}\Big)^{1/2} A_c^{1/3} \label{eq5}
\end{equation}
$R_0$ is the range of the effective nucleon potential taken as $1.4 fm$ and $\sigma$ is the droplet surface tension.  A temperature dependent formula for $\sigma$ is given in \cite{san}.  However, for the relatively low temperatures in this study, it can be treated as a constant $1.12 MeV fm^{-2}$. In nucleation theory $A_c$, the critical cluster size, is viewed as the size below which clusters break down and above which clusters grow. In our application of this approach, we treat both $\tau$ and $A_c$ as free parameters.  

There is some experimental information on fission times available from previous theoretical and experimental work. While calculated saddle to scission times are relatively short, of the order of $1000 fm/c$ \cite{negele}, some experiments have revealed significantly larger times for a large fraction of the events \cite{hofman, lazarev, molitoris, barrue}. Crystal blocking experiments find  long time components to the fission process that may extend out to $10^5 -10^7 fm/c$ \cite{barrue}.   

Once again after surveying a wide variety of parameter combinations, including temperature, density, proton fraction, time, and $A_c$ values we adopted a multi-parameter minimization technique to simultaneously fit the available experimental data. The Figure \ref{Fig2}b shows results of the fit in which the addition of the time dependence of the nucleation prevents the yields of heavier isotopes from achieving the NSE equilibrium values. The temperature, density and proton fraction fit parameters derived here are the ones used for  Figure \ref{Fig2}a. We see that addition of time $= 6400 fm/c$ and $A_c = 5.4$ to the fits provide a much better representation of the experimental yields. The fit metric,  $M^2= 1.18$ over the entire range of isotopes. For a fitting  range $Z\le 6$, that employed by Lestone in his paper, $M^2= 1.19$. 

Dynamic models based upon an evaporative approach invoke significant   barrier lowering in the neck region in attempts to explain both yields and energy spectra \cite{lestone, delion} of the light ternary particles. In the more microscopic work of Delion et al., the cluster emission is treated as decay from a resonant state in the neck region. In  the  model proposed here it has been assumed that,  at the time and density at which the cluster emission occurs,  the emission barrier has diminished to the point  where it is negligible. The suppression of the heavier cluster yields is attributed to nucleation time requirements.

As shown in this paper, the assumption of a nucleation modulated approach to nuclear statistical equilibrium, with reasonable parameters, provides a rather good fit to the ternary fission data. The success suggests that the process is dominated by cluster formation in low temperature low density nucleonic matter.  Naturally there is some interplay among the parameters and slight variations in one may be compensated for by changes in another. 

The present approach is useful in understanding some of the main features of the ternary fission data. For example the yield trend for $Z = 1$ and $Z=2$ yields is well reproduced and the absence of $^{3}$He can now be understood as reflecting the very large yield difference for the mirror nuclei $^{3}$H and  $^{3}$He. The former has a yield of $1.56\times 10^{-4}$ per fission while the latter is not detected. As equilibrium is approached the ratio of free neutrons to free protons (those not bound in clusters) becomes many times larger than the N/Z of the fissioning system. In the case considered here the calculation represented in Figure \ref{Fig2}b results in a free neutron to proton ratio of $4.97\times 10^3$. Such a ratio provides a natural explanation for the low scission proton yield and for missing $^{3}$He. At equilibrium the $^{3}$H/$^{3}$He ratio is directly related to the free $n/p$ ratio. Based on the observed $^{3}$H yield and using the calculated free neutron to proton ratio we predict a $^{3}$He yield of $1.81\times 10^{-8}$. This is apparently below the experimental detection limit. We note that similar reasoning leads to a $^{7}$Li/$^{7}$Be (also mirror nuclei) ratio of $1.61\times 10^4$. Based upon the reported $^{7}$Li yield this leads to a predicted $^{7}$Be yield of $8.22\times10^{-11}$. The experiment reports an upper limit $\le 3.96 \times 10^{-8}$. Other calculated values presented in Figure \ref{Fig2} for unobserved isotopes constitute very useful predictions for their yields. 

Free cluster binding energies are employed in this work and these are consistent with very low derived densities where the NSE approach is expected to be valid. However, it is possible that medium modifications of the binding energies at higher densities play a role in determining the cluster yields \cite{ropke, hagel1}. Ignoring this effect could lead to an error in the parameter determinations. A calculation including this medium modification is beyond the scope of the present work but would certainly be interesting. 

\begin{figure}
\centering
\includegraphics[width=0.5\columnwidth]{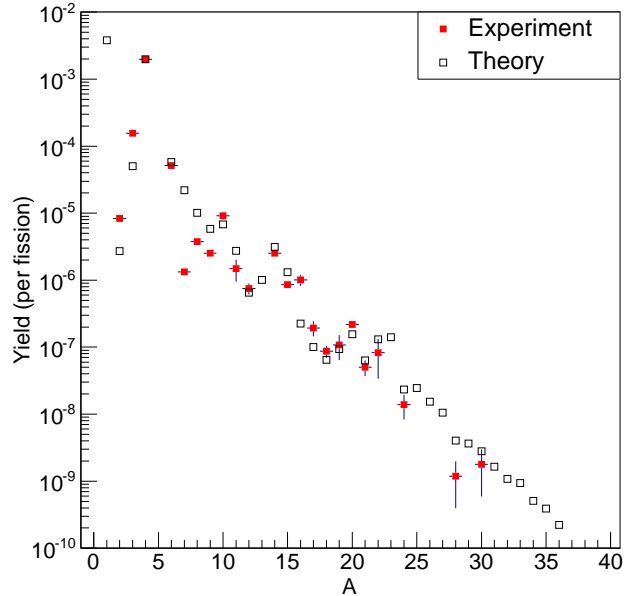}
\caption[]{(Color online) Yield per fission as a function of the fragment mass number (A).  Solid points represent $^{241}Pu(n_{th}, f)$ experimental yeilds from Koester {\it et al} \cite{koester1}.  Open data points are the result of NSE calculations \cite{heeg} with the same parameters featured in the bottom panel of Figure \ref{Fig2}. $M^2$ fit metric $= 0.561$.} \label{Fig3}
\end{figure}

In general nucleation treatments use simple scaling relationships to calculate the cluster formation free energies. In the approach proposed here for nuclear clusters experimental binding energies are used. As a result the isospin, coulomb, and pairing effects on the isotope formation free energies (and thus on the yields) are already introduced through use of these binding energies. The use of these and the associated isotope specific internal partition functions add a level of structural detail not intrinsic in nucleation treatments. Since the nucleation model we have employed makes no distinction between protons and neutrons, it is useful to ask whether fits to the isotope mass distributions make any significant change in the quality of the model fits. In Figure \ref{Fig3} we present results of a minimization procedure fitting the experimental mass distribution. Here we see that the quality of the fit to the observed mass distribution ( $M^2 = 0.561$ over the entire range of isotopes) is significantly better that of the fit to the isotopes presented in Figure \ref{Fig2}b. This suggests that a binary system nucleation approach, treating neutrons and protons separately, might offer some improvement in modeling the isotope yields and this should be investigated in future work.

Of the previous approaches advanced, the treatment by Valskii \cite{valskii} is most closely related to the present one.  Using four parameters , i.e.,  the proton work function, the neutron work function, temperature and a scale parameter and assuming all observed isotope yields to reflect statistical equilibrium, Valskii fit the yields of detected light isotopes and employed the resultant fit parameter values to predict yields for unobserved species. These predicted yields for heavier isotopes were far above the observed yields, a result similar to that presented in Figure \ref{Fig2}a in which the yields from the basic NSE model are shown.  
 
The results of the present study have several important implications. First, in the picture presented here, the low energy light particle or fragment accompanied ternary fission may be viewed as a manifestation of a ``pastina'' phase in the low density neck matter produced during nuclear fission. This might be considered as the mesoscopic analog of the nuclear pasta postulated in the skins of neutron stars or supernovae \cite{newton, watanabe, pais, schneider}. As discussed in a recent paper based on the use of energy density functionals to explore the cluster to liquid phase transition in nucleonic matter, larger pasta(or crystalline) structures are not expected in finite nuclei \cite{ebran}.

Second, clusterization effects such as those manifested here should not be restricted to fission necks but should be a general property of nuclei, which are leptodermous systems, and therefore clusterization in the lower density skin may modify the skin properties. Thus quantitative analyses of the data on skin thicknesses may require that the correlations which lead to the cluster formation be included. This could have an effect on analyses presently being carried out to extract information on the slope of the symmetry potential near normal density, for example \cite{usmani}. 

Third, the derived fit parameters imply that the light particle or fragment accompanied ternary fission occurs on a time scale which is long compared to calculated saddle to scission times and thus may belong to the very slow fission components revealed by several fission time measurements \cite{molitoris}.  It would be interesting to have experimental data on this point.  

Fourth, we note that data and analyses of the type considered here provide additional access to a low density, low temperature region of the nuclear phase diagram which can provide useful tests for modeling astrophysical phenomena \cite{fattoyev}. Analogous analyses of  neck emission particles from peripheral and semi-peripheral reactions could allow studying the process at higher temperatures and following the cluster formation process in greater detail. 

\section*{Acknowledgements}
This work was supported by the United States Department of Energy under Grant \#DE-FG03-93ER40773 and by The Robert A. Welch Foundation under Grant \#A0330.

\end{document}